# Experimental realization of graphene-like borophene

Wenbin Li[1,2], Longjuan Kong[1,2], Caiyun Chen[1,2], Jian Gou[1,2], Shaoxiang Sheng[1,2], Weifeng Zhang[3], Hui Li[4], Lan Chen[1,2*], Peng Cheng[1,2*], and Kehui Wu[1,2,5*]

[1]*Institute of Physics, Chinese Academy of Sciences, Beijing 100190, China*
[2]*School of physical sciences, University of Chinese Academy of Sciences, Beijing 100049, China*
[3]*School of Physics and Electronics, Henan University, Kaifeng 475004, Henan Province, China*
[4]*Beijing Advanced Innovation Center for Soft Matter Science and Engineering, Beijing University of Chemical Technology, Beijing 100029, China*
[5]*Collaborative Innovation Center of Quantum Matter, Beijing 100871, China*
*e-mail：lchen@iphy.ac.cn, pcheng@iphy.ac.cn, khwu@iphy.ac.cn

***Abstract***

We report a successful preparation of a purely honeycomb, graphene-like borophene, by using an Al(111) surface as the substrate and molecular beam epitaxy (MBE) growth in ultrahigh vacuum. Our scanning tunneling microscopy (STM) reveals perfect monolayer borophene with planar, non-buckled honeycomb lattice similar as graphene. Theoretical calculations show that the honeycomb borophene on Al(111) is energetically stable. Remarkably, nearly one electron charge is transferred to each boron atom from the Al(111) substrate and stabilizes the honeycomb borophene structure, in contrast to the little charge transfer in borophene/Ag(111) case. This work demonstrates that the borophene lattice can be manipulated by controlling the charge transfer between the substrate and the borophene film.

*Introduction*

The boom of graphene research has inspired the theoretical prediction and experimental discovery of a number of elemental two-dimensional (2D) materials, such as silicene[1-3], germanene[4-6], stanene[7, 8] and phosphorene[9]. Different from the planar honeycomb structure with $sp^2$ hybridization in graphene, these 2D materials tend to form buckled honeycomb lattice with mixed $sp^2$-$sp^3$ hybridization, due to their larger atomic radius as compared with carbon. An exceptional case is borophene, since boron possesses an even smaller atomic radius than carbon. However, as boron has only three valence electrons, the electron deficiency makes a honeycomb lattice unstable in boron. Instead, a mixture of honeycomb units together with triangular units was predicted to be more stable[10-12]. Various stable structures of borophene sheets were predicted, like α, β and so on, with different arrangements of the honeycomb and triangular units[10, 11, 13-19]. And a few of them have been successfully synthesized on silver surface[20-24].

With mixed honeycomb and triangular units, these borophene lattices can be regard as a triangular lattice with periodic holes, or inversely, as a honeycomb lattice with periodic boron adatoms. Therefore, a challenging question is that whether it is possible to prepare a borophene monolayer with a pure honeycomb lattice, or in other words, a graphene-like borophene. A honeycomb borophene is important due to two reasons. Firstly a honeycomb lattice will naturally host Dirac fermions and thus intriguing electronic properties like other as group IV elemental 2D materials[25]. Secondly, in the well-known high Tc superconductor, $MgB_2$, the crystal structure consists of boron planes and intercalated Mg atoms, where the boron plane has a pure honeycomb structure like graphene[26]. It is remarkable that in $MgB_2$, superconductivity occurs in the boron planes, while the Mg atoms serves as electron donors[26]. Therefore, preparation of a honeycomb 2D boron lattice may open up opportunities in controlling the high Tc superconductivity by tuning the structure of boron-based compounds[27-29].

In this work, we report a successful preparation of a purely honeycomb, graphene-like borophene, by using an Al(111) surface as the substrate and molecular beam epitaxy (MBE) growth in ultrahigh vacuum. The idea is that aluminum has three free electrons and thus can provide an effective compensation for the electron deficiency in borophene. Our scanning tunneling microscopy (STM) reveals perfect monolayer borophene film with honeycomb lattice. On the other hand, theoretical calculations show that the honeycomb borophene on Al(111) is energetically stable. Remarkably, nearly one electron charge is transferred to each boron atom from the Al(111) substrate, in contrast to the little charge transfer in B/Ag(111) case. This work demonstrates that the borophene lattice can be manipulated by controlling the charge transfer between the substrate and the borophene film[30, 31]. And the successful fabrication of honeycomb borophene provides attractive possibility to control superconductivity in boron-based compounds.

*Methods*

The experiments were performed both in a homebuilt low temperature-STM-MBE system and a Unisoko USM 1300 system. Clean Al(111) substrate was prepared by cycles of $Ar^+$ ion sputtering and annealing. The borophene monolayers were prepared by evaporating pure boron (99.999%) on Al(111) with the substrate temperature at about 500 K. STM and scanning tunneling spectroscopy (STS) data were obtained either at 77 K or 4.5 K. The STM data were processed using the free WSxM software[32]. First principles calculations were performed within the framework of plane-wave density functional theory (DFT), as implemented in the Vienna *ab initio* Simulation Package (VASP)[33]. Generalized gradient approximation (GGA) with Perdew–Burke–Ernzerhof (PBE) function[34] was adopted to describe the exchange-correlation interaction. The interaction between valence and core electrons was described by the projector-augmented wave (PAW)[35]. For geometric optimization, both lattice constants and positions of atoms are relaxed until the force on each atom is less than 0.001 eV/ Å and the criterion for total energy convergence is $1.0 \times 10^{-5}$ eV/atom. A kinetic energy cutoff of 400 eV is chosen for the plane-wave expansion. We applied periodic boundary conditions in all directions with a vacuum layer larger than 15 Å to avoid image-image interaction along the sheet thickness. The Brillouin zone was sampled by a 21 × 21 ×1 Monkhorst-Pack *k*-mesh[36].

*Results and Discussion*

Monolayer borophene islands are formed upon evaporation of boron on clean Al(111) substrate at about 500 K, as shown in Fig. 1(a) and (b). The measured height profile in Fig. 1(c) shows a single Al(111) step with height of 236 pm, as well as a height difference between monolayer borophene and the Al(111) substrate (320 pm). It is notable that the height of monolayer borophene exhibits little change with varying the scanning bias between (-6 V, +6 V) (as shown in Fig. 1(d)). With the increase of boron coverage, the borophene islands grow in size, and some of them can cross the Al(111) substrate steps, as shown in Fig. 1(b) for example. The height profile in Fig. 1(e) reveals a single Ag(111) step, with height of 230 pm, underneath the island without breaking the continuity of the borophene lattice.

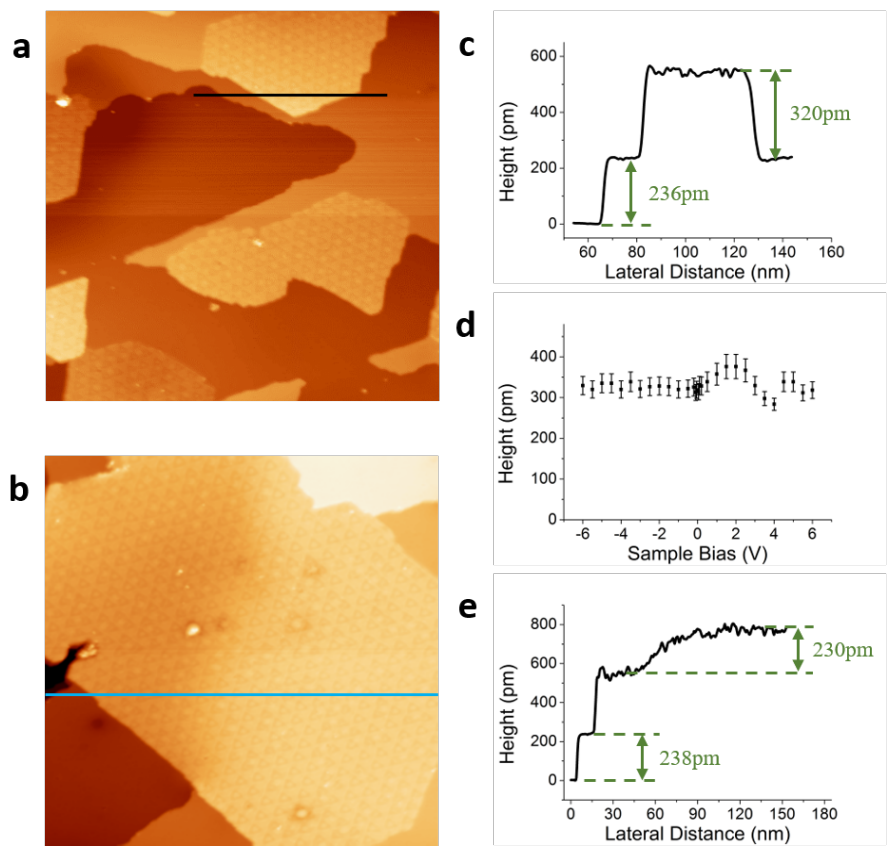

**Figure 1**. Monolayer borophene on Al(111). (a) STM image (180 nm × 180 nm) of borophene islands on Al(111) surface; (b) STM image (150 nm × 150 nm) showing a monolayer borophene island running across an Al(111) step. (c) Line profile along the black line in (a). (d) the measured height of single layer borophene at different bias voltages shows little change. (e) Line profile along the blue line in (b). The scanning parameters are: (a) sample bias -4.0 V, I = 50 pA; (b) sample bias -2.0 V, I = 50 pA.

High-resolution STM images reveal characteristic quasi-periodic, triangular corrugations on the surface of the borophene monolayer, as shown in Fig. 1 and the zoom-in image in Fig. 2(a). The period of the corrugation is roughly 7 nm and height difference in the z directions are 40 - 60 pm. Such large corrugation pattern often occurs due to strain relaxation on the surface, such as the herringbone pattern on Au(111) surface[37, 38], as will be discussed later. Importantly, atomically resolved STM images show a honeycomb lattice structure, as shown in Fig. 2(b). The measured lattice constant is 0.29 nm, which is bit smaller than the calculated lattice constant of a free standing, honeycomb borophene (0.3 nm), and more close to the lattice constant of Al(111)-1×1 (0.286 nm). The honeycomb lattice is locally flat and without buckling, since the atoms in A and B sites of the honeycomb lattice appears equivalent in the STM images in Fig. 2(b). The honeycomb lattice is continuously overlapped on the large-periodic, triangular corrugation, as clearly illustrated in the 3D STM image in Fig. 2(c), even though the atomic corrugation in the honeycomb lattice (~1.5 pm) is much smaller than the triangular corrugation (40 - 60 pm). To conclude, our STM data show unambiguously that we have obtained honeycomb, graphene-like borophene monolayer on Al(111).

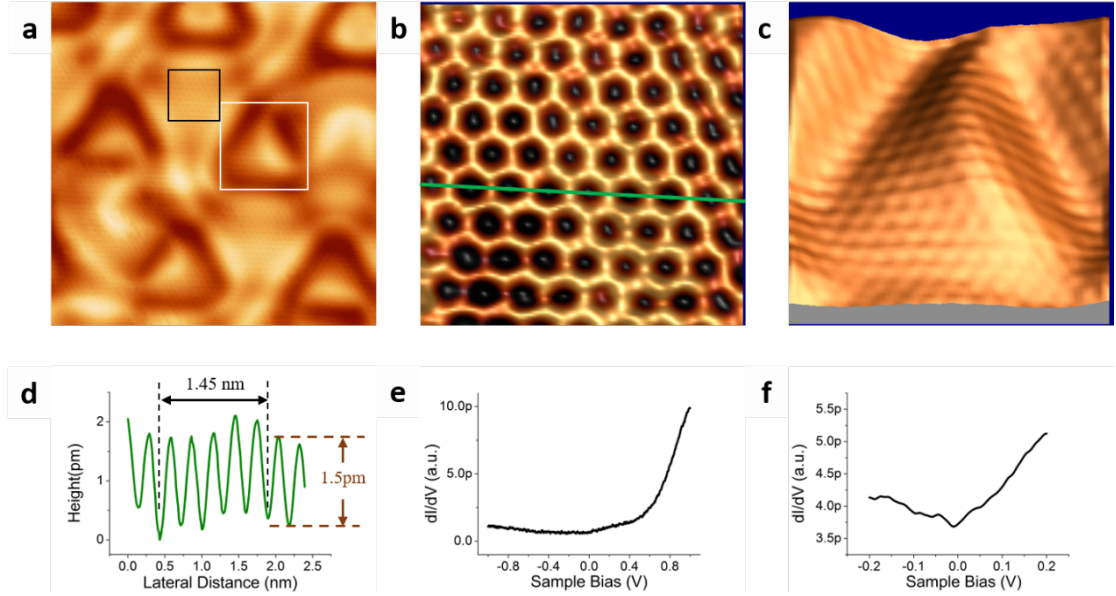

**Figure 2**. High resolution STM images and STS data on borophene monolayer on Al(111). (a) STM image (15 nm × 15 nm) showing the large-period, triangular corrugation. (b) A high resolution STM image (2.4 nm × 2.4 nm) of the area marked by black rectangle in (a), showing a flat honeycomb lattice. (c) 3D STM image (4 nm × 4 nm) of the area marked by white rectangle in (a). (d) line profile along green line in (b). (e, f) dI/dV curves taken on borophene surface with different bias voltage range. The scanning parameters for (a-c) are: tip bias 11 mV, I = 130 pA.

Previous theoretical works have already indicated that the honeycomb lattice is unstable for free-standing borophene[10]. Obviously, the Al(111) substrate play the crucial role in stabilizing the honeycomb borophene lattice. To understand the mechanism in detail, we performed first principle calculations and compared the stability of borophene on Al(111) and Ag(111), respectively. Honeycomb borophene 1×1 monolayer was overlapped with 1×1 unit cells of Al (111) and Ag (111) surfaces, respectively, as these two surfaces both have very small lattice mismatch with honeycomb borophene 1×1. The substrates consist of 4 metal slabs with the bottom two slabs fixed, while the other two slabs and the borophene sheet fully relaxed in the geometric optimization. To estimate the stability of the hexagonal borophene sheet on different metal substrates, the formation energy is defined as:

$$E_f = (E_{tot} - E_{sub} - N \times E_B)/N,$$

where $E_{tot}$ is the total energy of monolayer borophene on the metal substrate, $E_{sub}$ is the total energy of the substrate, $E_B$ is the energy per atom in the solid boron of α-$B_{12}$ phase, and N is the number of boron atoms in each unit cell[39]. Under such definitions, a configuration with smaller $E_f$ is more stable. After relaxation, the borophene lattice adopts a 1×1 lattice matching with the Al(111) substrate, with boron atoms located at the hollow site of the Al(111) lattice (Fig. 3(a)). The side view of the structure indicates that the borophene lattice is flat (as shown in Fig. 3(b)). Remarkably, there is a large amount of electron charge located at the interface between the borophene monolayer and the Al(111) substrate, as revealed in the charge distribution map (Fig. 3(c)), suggesting a pronounced electron transfer from the first layer of the substrate to the borophene monolayer. Meanwhile, the Bader charge analysis quantify that, in average, about 0.7 e electron is transferred to each boron atom from the Al substrate. Similar theoretical calculations were also performed on Ag(111) surface (Fig. 3(d-f)). However, the formation energy of honeycomb borophene on Ag(111) is much bigger than that on Al(111), and also much bigger than other reported borophene structures ( $β_{12}$ and $χ_3$ phases). This is consistent with previous conclusions that the honeycomb borophene on Ag(111) is unstable. In addition, there is little electron accumulation at the interface, indicating negligible charge transfer between borophene and the Ag(111) substrate. The average formation energy and charge transfer of honeycomb borophene on Al(111) and Ag(111) substrates were list in Table 1.

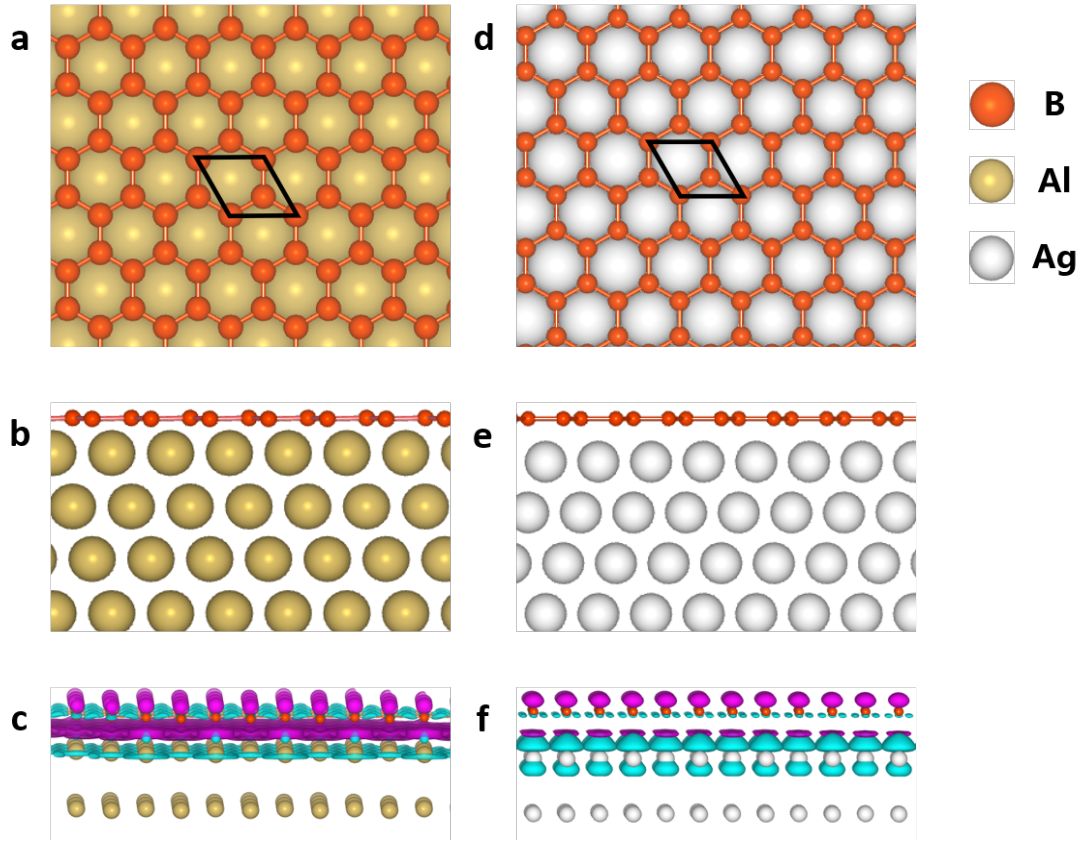

**Figure 3.** Structural models and electron density of honeycomb borophene on Al(111) and Ag(111). (a, b) top and side views of honeycomb borophene on Al(111). (c) electron density map of honeycomb borophene on Al(111). (d, e) top and side views of honeycomb borophene on Ag(111). and (f) is the corresponding electron density map. In (c) and (f), the electron accumulation and depletion regions were presented by red and blue colors, respectively.

**Table 1.** Formation energy and averaged charge transfer in different structures of borophene. h-BS and $\beta_{12}$-BS represent honeycomb and $\beta_{12}$ borophene structure, respectively.

|  | $E_f$ (eV/atom) | Average charge per boron \|e\| |
|---|---|---|
| h-BS on Al (111) | 0.31 | -0.70 |
| h-BS on Ag (111) | 0.81 | -0.06 |
| $\beta_{12}$-BS on Ag(111) | 0.35 | -0.03 |

It is not difficult to understand why Al(111) substrate can stabilize the honeycomb borophene lattice. Free-standing, honeycomb borophene lattice is unstable because the electron deficiency of boron atoms. Thus, boron adatoms are required to fill in some hexagonal units to compensate for the electron deficiency[40]. On the other hand, if the substrate can provide one electron to each boron atom, the boron atoms would become carbon-like, and forms a honeycomb lattice just like graphene. Aluminum is the best candidate for this purpose because it has three free electrons per Al atom.

A graphene-like 2D lattice will naturally result in the unique Dirac bands in its electronic band structure. The honeycomb borophene monolayer on Al(111) should possess the same in-plane band structure as a free-standing, honeycomb borophene, with possibly a shift of Fermi level due to the electron transfer from the substrate[41]. The local electron states of honeycomb borophene were studied by STS as shown in Fig. 2(e). The dI/dV spectroscope indicate the honeycomb borophene is metallic. In particular, a V-shape dI/dV curve is revealed near the Fermi energy, with a dip at the Fermi energy (Fig. 2(f)), such feature is characteristic to existence of the Dirac bands, similar as in graphene.

Finally, we discuss the mechanism for the formation of the quasi-periodic, triangular corrugation patterns. The amplitude of the triangular corrugation is about 40 to 60 pm and the distance between two triangles ranges from 6 nm to 7.5 nm. We noticed that the lattice constant of the honeycomb borophene lattice is varied slightly relative to the triangular corrugations. The periods are close to 0.29 nm in the flat area, and close to 0.3 nm in corrugated area. Regarding the fact that the borophene lattice is slightly compressed due to the interaction with the substrate, the appearance of such large period corrugation is very similar to the herringbone structure on Au(111) surface, where the relaxation of the compressed Au(111) lattice results in alternative hcp and fcc domains in a $22 \times \sqrt{3}$ superstructure. However, more detail investigation on the triangular patterns requires heavy calculation task and is still under progress.

*Conclusions*

We have successfully synthesized graphene-like, flat honeycomb borophene monolayer on Al(111). Theoretical calculations found that the honeycomb borophene is stable on Al(111) surface, and there is nearly one electron transferred from the Al(111) substrate to each boron atoms, which is the key for stabilizing the structure. Our work vividly demonstrated that one can manipulate the borophene lattice by controlling the charge transfer between the substrate and the borophene. And the honeycomb borophene provides attractive possibility to control superconductivity in boron-based compounds.

*Acknowledgement*

This work was supported by the National Key R&D Program under Grant No. 2016YFA0300904 and 2016YFA0202301the National Natural Science Foundation of China under Grant No. 11334011, 11674366 and 11674368, and the Strategic Priority Research Program of the Chinese Academy of Sciences under Grants No. XDB07010200 and XDPB06.